\documentstyle[twoside,fleqn,espcrc2]{article}

\newcommand{\AmS}{{\protect\the\textfont2
  A\kern-.1667em\lower.5ex\hbox{M}\kern-.125emS}}

\title{The $\rho $ Meson and the Thermal Behavior of an Effective Hadronic
Coupling Constant}

\author{
{\bf C. A. Dominguez}$^{a}$, {\bf M. S. Fetea}$^{a}$, and
 {\bf M. Loewe}$^{b}$\\
\vspace{.5cm}
$a$ Institute of Theoretical Physics and Astrophysics,\\ 
    \mbox{} \hspace{.15cm} University of Cape Town, Rondebosch 7700, South 
    Africa\\
$b$ Facultad de Fisica,
    Pontificia Universidad Catolica de Chile, Santiago, Chile
}
      
\begin{document}

\begin{abstract}
Vector Meson Dominance ideas together with a Finite Energy QCD sum rule
allows for the determination of the $q^{2}$- and the $T$- dependence of the effective
hadronic coupling constant $g_{\rho \pi \pi}$ in the space-like region. 
It turns out that $g_{\rho \pi \pi}(q^{2},T)$ vanishes at the critical 
temperature $T_{c}$, independently of $q^{2}$. A comparison with a previous 
independent QCD determination of the electromagnetic pion form factor at finite
temperature supports the validity of Vector Meson Dominance at finite temperature.
We find also thet the pion radius increases with $T$, having a divergent behavior
at $T_{c}$.
\end{abstract}

\maketitle

In this talk \footnote{talk given at the QCD Euroconference 97, Montpellier 
3-9 July 1997}
 we will consider the thermal behavior of the effective hadronic
coupling $g_{\rho \pi \pi}(q^{2},T)$. From the phenomenolgical point of view,
this is an important issue since this coupling is related to the electromagnetic
pion form factor which plays an important role in the production
of dileptons, in relativistic heavy ion collisions,
due to pion annhilation in the normal hadronic phase.\cite{FK}-\cite{DL1}.

\smallskip
To answer this question we will use Finite Energy QCD sum rules (FESR)
at finite T for the three point function that involves the rho-meson
interpolating current plus two axial divergences. We will start first with the
determination of $g_{\rho \pi \pi}(q^{2})$ at zero temperature (see \cite{NP}
for an earlier determination using Laplace Sum Rules). In this way we will 
establish the normalizations, checking also the 
validity of Vector Meson Dominance (VMD) here. 
This will be done through a comparison with an independent determination
of the electromagnetic pion form factor, not relying on VMD.
Since the latter does fit the data at $T=0$ very well, we can adopt it as
the benchmark $F_{\pi }(Q^{2},T)$ at finite T.

\bigskip
Later we will compare our result 
at finite temperature with the same direct determination \cite{DLR}, finding
an agreement  between the two expressions which can be taken as 
evidence in support of VMD at finite T.

Let us consider the $T=0$ correlator

\begin{eqnarray}
\Pi_{\mu }(q) &=& i^{2}\int \int d^{4}x d^{4}y e^{-iq\cdot y} e^{ip'\cdot x} \nonumber \\
& &< 0| T (j_{\pi}^{\dag} (x) J_{\mu}^{\rho} (y) j_{\pi} (0) |0> \nonumber \\
&=& \Pi_{1} (q^{2}) \; P_{\mu} + \Pi_{2} (q^{2}) \; q_{\mu}
\end{eqnarray}

\noindent
where $J^{\rho}_{\mu}(y)$ and $j_{\pi} (x) $ are the usual $\rho $- and $\pi $- 
meson currents. $q_{\mu} = (p' - p)_{\mu}$ 
and $P_{\mu} = (p' + p)_{\mu}$.

To leading order in $\alpha _{s}$ and in the quark masses, the imaginary part
of this three point function comes only from the perturbative triangle diagram
and 
is given by

\begin{eqnarray}
Im\Pi_{\mu }|_{QCD} = \frac{3}{4} \frac{(m_{u} + m_{d})^{2}}{[(s + s' +Q^{2})^{2} 
 -4ss']^{\frac{3}{2}}}\nonumber\\
 \mbox{} \times [-Q^{2}ss'P_{\mu } + ss'(s-s')q_{\mu }]
\end{eqnarray}

\noindent
where $s = p^{2}$, $s' = p'^{2}$, and $Q^{2} = -q^{2} \geq 0$. For 
the hadronic part of the analysis we saturate with a pion intermediate state and 
use the current-field
identity $j_{\mu }^{a} =\frac{MÐ{\rho }^{2}}{f_{\rho }} \rho_{\mu }^{a}$, beeing a
an isospin index and $\rho_{\mu }^{a}$ the rho-meson field. Experimentally 
$f_{\rho } = 5.0 \pm 0.1$.

\pagebreak
For the hadronic spectral function, e.g $Im \Pi_{1}$ we get
\begin{eqnarray}
Im\Pi_{1}(s,s',Q^{2})|_{HAD}  =  -2f_{\pi }^{2}\mu_{\pi }^{4} \frac{M_{\rho }^{2}}
{M_{\rho }^{2} + Q^{2}}\nonumber\\ 
\mbox{} \times \frac{g_{\rho \pi \pi }(Q^{2})}{f_{\rho }}\pi ^{2}
\delta (s - \mu_{\pi }^{2})\delta(s' - \mu_{\pi }^{2})\nonumber\\
\mbox{} + \theta (s-s_{0})\theta (s' -s'_{0})Im\Pi_{1}(s,s',Q^{2})|_{QCD} 
\end{eqnarray}

In the above expression $f_{\pi } = 93.2 MeV$ and 
$ g_{\rho \pi \pi }(M_{\rho }^{2}) = 6.06 \pm 0.03$.
Note that, according to the philosophy of QCD sum rules, we assume that
for s and \'s values above the continuum thresholds the hadronic spectral function 
can be expressed in terms of QCD degrees of freedom. Due to the radial
excitations
of the rho-meson, $g_{\rho \pi \pi }$ actually is a form factor, i.e. a function
of $Q^{2}$. In the dual model \cite{dual} this effect is taken into account. Note that in
addition to the pion pole there are other terms, as for example the $a_{1}$ meson. Here,
however, we shall include these in the hadronic continuum provided that the threshods $s_{0} 
\simeq s'_{0} > 1 - 3$ $GeV^{2}$. Now we can proceed in the standard fashion, i.e.,
invoking Cauchy's theorem we construct
the lowest dimensional FESR 
\begin{eqnarray}
\int_{0}^{s_{0}}\int_{0}^{s'_{0}}Im \Pi _{1}(s, s')|_{HAD}ds ds'=\nonumber\\
\; \; \; \; \; \mbox{} = \int_{0}^{s_{0}}\int_{0}^{s'_{0}}Im \Pi _{1}(s, s')|_{QCD}ds ds'
\end{eqnarray}

\noindent
where $s_{0}$ and $s'_{0}$ are the continuum thresholds, i.e. the onset of perturbative
QCD. In this way we get

\begin{equation}
\frac{g_{\rho \pi \pi }(Q^{2})}{f_{\rho }} = \frac{3}{8\pi ^{2}}\frac{f_{\pi }^{2}}
{<\overline{q} q>^{2}}\frac{Q^{2}}{M_{\rho }^{2}} (Q^{2} + M_{\rho }^{2})I(q^{2})
\end{equation}

\noindent
where
\begin{equation}
I(Q^{2}) = \frac{s_{0}}{16}(3 + \frac{s_{0}}{Q^{2}}) + 
\frac{1}{8}(s_{0} + \frac{3}{4}Q^{2} ) \ln{\frac{Q^{2}}{Q^{2} + 2s_{0}}}
\end{equation}

We have used the Gell-Mann, Oakes and Renner (GMOR) relation \cite{GMOR}.The 
result for $I(Q^{2})$ was obtained after a double integration 
in a triangle in the s, \'s plane.
It is important to remark that
other shapes for the integration region do not introduce appreciable differences in the 
numerical results. According to Extended Vector Meson Dominance (EVMD),
where a $Q^{2}$-dependence
of the effective coupling $g_{\rho \pi \pi }$ is allowed, 
the electromagnetic pion form is given by

\begin{equation}
F_{\pi }(q^{2})|_{EVMD} = \frac{M_{\rho }^{2}}{M_{\rho }^{2} + Q^{2}}
\frac{g_{\rho \pi \pi }(Q^{2})}{f_{\rho }}
\end{equation}

\noindent
which, after substituting the result from the FESR leads us to

\begin{equation}
 F_{\pi }(Q^{2})|_{EVMD}= \frac{3}{8\pi ^{2}}\frac{f_{\pi }^{2}}
{<\overline{q} q>^{2}}Q^{2} I(q^{2})
\end{equation}

\noindent 
where $I(Q^{2})$ was defined in Eq.(6). This result should be compared with a
previous independent determination of $F_{\pi }$ based on a three-point function involving
the electromagnetic current and two axial currents, \cite{russians}, which projects the
form factor directly, making no use of VMD, i.e.

\begin{equation}
F_{\pi }(Q^{2}) = \frac{1}{16 \pi ^{2}} \frac{1}{f_{\pi }^{2}} \frac{s'_{0}}
{(1 + Q^{2}/2s'_{0})^{2}}
\end{equation}

It is remarkable that Eqs. (8) and (9), in spite of looking structurally very different,
are numerically very similar for $Q^{2} \geq 0.5 \: GeV^{2}$, if we 
take $s_{0} \simeq 2.18 \: GeV^{2}$, and  $s'_{0} \simeq 1 \: GeV^{2}$. The latter values 
leads to a very good
fit of the data above $1 \:GeV^{2}$. It should be stressed, since the correlators
are different, that the onset of the continuum need not be the same in both cases. All one knows
is that $s_{0} (s'_{0})$ should roughly be in a region where the resonances
loose prominence, and the hadronic continuum takes over. Somewhere between $1-3 \: GeV^{2}$.
The good numerical agreement between both results should be considered as a reflection of 
the validity of EVMD. 

\smallskip
Our purpose here, however, is not another fit to the data at $T=0$ but rather
the thermal evolution of the $\rho \pi \pi $ form factor. Therefore, let us reconsider the FESR
at finite temperature. The thermal corrections to $\Pi _{1}(Q^{2})|_{QCD}$
can be calculated using the
finite temperature fermion propagators. Since this is a one loop calculation we do not
need the full machinery of Thermo Field Dynamics. For the imaginary part we find

\begin{eqnarray}
Im \Pi_{1}(s, s', Q^{2}, T) & = & Im\Pi_{1}(s, s', Q^{2}, 0)\nonumber\\
& & \mbox{} \times F(s, s', Q^{2}, T)
\end{eqnarray}

\noindent
where

\begin{eqnarray}
F(s, s', Q^{2}) & = & 1 - n_{1} - n_{2} - n_{3}\nonumber\\
& & \mbox{} + n_{1}n_{2} + n_{1}n_{3} + n_{2}n_{3}
\end{eqnarray}

\begin{equation}
n_{1} \equiv n_{2} \equiv n_{F}( \vert \frac{1}{2T}\sqrt{\frac{x+y}{2}} \vert)
\end{equation}

\begin{equation}
n_{3} \equiv n_{F}( \vert\frac{Q^{2} + (x-y)/2}{2T\sqrt{\frac{x+y}{2}}} \vert)
\end{equation}

\noindent
$n_{F}(x) = (1+e^{x})^{-1}$ is the Fermi-Dirac factor and $x=s+s'$, $y=s-s'$. Concerning
the hadronic side, both $f_{\pi }$ and $<\overline{q} q>$ will develope a temperature
dependence, and so will $S_{0}$. We can safely
ignore the small variations of $M_{\rho }(T)$ from QCD sum rules
\cite{CAD} and other methods \cite{DLRoj}.
The temperature behavior of the asymptotic freedom threshold
can be obtained from an independent FESR associated to a two-point function correlator
involving 
axial-vextor currents \cite{threshold1} \cite{threshold2}. 
Here $f_{\pi }(T)$ is an input. It turns out that for
temperatures not too close to the critical temperature $T_{c}$, the following scaling relation
holds to a very good approximation \cite{dfl}.

\begin{equation}
\frac{f_{\pi }^{2}(T)}{f_{\pi }^{2}(0)} \simeq  \frac{<\overline{q} q>_{T}}
{<\overline{q} q>} \simeq \frac{s_{0}(T)}{s_{0}(0)}
\end{equation}

We will make use of the results of \cite{Bar} for $f_{\pi }(T)$ and $<\overline{q} q>_{T}$
in the chiral limit and also for $m_{q} \neq 0$. In addition we invoke the GMOR relation at
finite T. It only gets modified at next to leading order in the quark masses \cite{dfl}. For
a different recent analysis of the same subject, with essentially the same numerical behavior
see also \cite{toublan}.

\bigskip
The $T \neq 0$ FESR now reads

\begin{eqnarray}
\frac{g_{\rho \pi \pi }(Q^{2},T)}{f_{\rho }} & = & \frac{3}{8\pi ^{2}}\frac{f_{\pi }^{2}(T)}
{<\overline{q} q>^{2}_{T}}\frac{Q^{2}}{M_{\rho }^{2}} (Q^{2} + M_{\rho }^{2})\nonumber\\
& & \mbox{} \times I(q^{2},T)
\end{eqnarray}

\pagebreak
\noindent
where now

\begin{eqnarray}
I(Q^{2},T) & = & \frac{1}{8}\int_{0}^{s_{0}(T)}
\!\!\!\!\!\!\! dx \int_{-x}^{x} \!\!\!\! dy\nonumber\\
& & \mbox{} \times \frac{(x^{2} - y^{2})}
{Q^{4} + 2xQ^{2} + y^{2})^{3/2}}\nonumber\\
& & \mbox{} \times F(x, y, Q^{2}, T)
\end{eqnarray}

\noindent
and the integration in Eq.(16) must be done numerically. In Fig.(1) we show the behavior
of the ratio $R_{1} \equiv g_{\rho \pi \pi }(Q^{2},T)/ g_{\rho \pi \pi }(Q^{2},0)$ 
for $f_{\pi }(T)$ and 
$<\overline{q} q>_{T}$ in the chiral limit (curve (a)) as well as for $m_{q} \neq 0$
(curve(b)) for $Q^{2} = 1 \:GeV^{2}$. Higher values of $Q^{2}$ give similar results.

\begin{figure}[htb]
\vspace{43 mm}
\caption{}
\label{fig:largenenough}
\end{figure}

It is 
important to remark that the vanishing of $R_{1}$ at or near the critical temperature
is basically $Q^{2}$ - independent, providing analytical evidence for deconfinement. The
good agreement between the pion form factor using $g_{\rho \pi \pi }(Q^{2})$ plus VMD
and that obtained without invoking VMD, persists at finite temperature.

\smallskip
Finally, an extrapolation to $Q^{2} = 0$ allows for a determination of the $\rho \pi \pi $
root mean squared radius. The ratio $<\! r^{2}\!>(T)/<\!r^{2}\!>(0)$ is well 
defined,
in spite of the fact that $<\!r^{2}\!>$ is divergent at any temperature due 
to mass singularities.
It turns out that this ratio increases with increasing T until the critical temperature where
it diverges, as we could have expected
from an intuitive point of view, 
thus signalling deconfinement. See \cite{dfl2} for more details.

\medskip
\noindent
\bf Acknowledgements: \rm The work of (CAD) and (MSF) has been supported in part by the FRD
(South Africa), and that of (ML) by Fondecyt (Chile) under grant No. 1950797, and
Fundaci\'on Andes under grant No. C-12999/2.

\noindent
\Large Discussions

\normalsize

\medskip
\noindent
\bf H. G. Dosch, \rm Inst. Theor. Phys., Heidelberg.

\noindent
\it Can you comment on the objections of B. Ioffe and coworkers that one should not use 
thermal quark propagators, but rather take into account the pionic heath bath?

\smallskip
\noindent
\bf M. Loewe

\noindent
\it We have investigated the validity of the thermal
OPE in the framework of two exactly solvable models: the two-dimensional $\sigma $-model $O(N)$, 
in the
large N limit, and the Schwinger model, showing that the thermal dependence of
the perturbative part cannot be absorbed into the condensates. 
No confusion arises in these cases. This point
strongly supports the thermal QCD sum rules program. It would be extremely bizarre
if duality and the OPE suddenly would be no longer valid, as soon as we increase 
the temperature in a few millikelvins. 
Both descriptions are complementary instead of beeing contradictory.
The pion basis is well suited to determine the temperature dependence 
of the vacuum condensates at low T. It does not make use of QCD-hadron duality. 
The quark-gluon basis, on the other hand, allows for an extension of the QCD sum rule
program.

\medskip
\noindent
\bf M. Neubert, \rm CERN

\noindent
\it The previous speaker (S. Mallik) has shown that for finite T new additional condensates
enter in the QCD sum rules. Would these effects have an impact on your analysis ?

\smallskip
\noindent
\bf M. Loewe

\noindent
\it In fact we have not used these new condensates. 
Remember thar for FESR, because of the dimensions involved, not all possible new condensates
will contribute to a particular FESR. 
However we have checked that the new condensates are negligible 
from the numerical point of view.


\begin{thebibliography}{9}
\bibitem{FK} For a rewiev see e.g. F. Karsch, Nucl. Phys. A 590 (1995) 367.
\bibitem{DL1} C. A. Dominguez and M. Loewe, Z. Phys. C, Particles \& Fields, 49 (1991) 423.
\bibitem{NP} S. Narison and N. Paver, Z. Phys. C, Particles \& Fields, 22 (1984) 69.
\bibitem{DLR} C. A. Dominguez, M. Loewe, and J. S. Rozowsky, Phys. Lett. B 335 (1994) 506.
\bibitem{dual} C. A. Dominguez, Phys. Rev. D 25 (1982) 3084; Mod. Phys. Lett. A 2 (1987) 983.
\bibitem{GMOR} M. Gell-Mann, R. Oakes, and B. Renner, Phys. Rev. 175 (1968) 2195.
\bibitem{russians} B. L. Ioffe and A. V. Smilga, Nucl. Phys. B 216 (1883) 373; V. A. 
Nesterenko and A. V. Radyushkin, Phys. Lett. B 115 (1982) 410; V. L. Eletsky and Y. I.
Kogan, Z. Phys. C, Particles \& Fields, 20 (1983) 357.
\bibitem{CAD} C. A. Dominguez, UCT report No. UCT-TP-218/94, and hep-ph/9410363 (1984).
\bibitem{DLRoj} C. A. Dominguez, M. Loewe and J. C. Rojas, Z. Phys. C, 
Particles \& Fields, 59 (1993) 63.
\bibitem{threshold1} C. A. Dominguez and M. Loewe, Phys. Lett B 233 (1989) 201.
\bibitem{threshold2} A. Barducci, R. Casalbuoni, S. de Curtis, R. Gatto, and G. Pettini,
Phys. Rev. D 46 (1992) 2203. 
\bibitem{dfl} C. A. Dominguez, M. S. Fetea, and M. Loewe, Phys. Lett. B 387 (1996) 151 and
Nucl. Phys. B (Proc. Suppl.) 54A (1997) 333.
\bibitem{Bar} A. Barducci, R. Casalbuoni, S. de Curtis, R. Gatto, and G. Pettini,
Phys. Lett. B 244 (1990) 311.
\bibitem{toublan} D. Toublan, Preprint Bern BUTP-97/14, hep-ph/9706273.
\bibitem{dfl2} C. A. Dominguez, M. S. Fetea, and M. Loewe, Phys. Lett. B 406 (1997) 149. 
\end{thebibliography}
\end{document}